# From Static to Dynamic: Evaluating the Perceptual Impact of Dynamic Elements in Urban Scenes via MLLM-Guided Generative Inpainting

Zhiwei WEI [a,b], Mengzi ZHANG [a], Boyan LU [a], Zhitao DENG [c], Nai YANG [d], Hua LIAO [a,b*]

[a] *Hunan Normal University, School of Geographic Sciences, Hunan Changsha, China;* [b] *Hunan Key Laboratory of Geospatial Big Data Mining and Application, Hunan Changsha, China;* [c] *Advanced Interdisciplinary Institute of Satellite Applications, State Key Laboratory of Earth Surface Processes and Hazards Risk Governance, Faculty of Geographical Science, Beijing Normal University, Beijing, China;* [d] *School of Geography and Information Engineering, China University of Geosciences, Wuhan, China*.

Address for correspondence: Hua Liao. E-mail: liaohua@hunnu.edu.cn

# From Static to Dynamic: Evaluating the Perceptual Impact of Dynamic Elements in Urban Scenes via MLLM-Guided Generative Inpainting

**Abstract:** Understanding urban perception from street-view imagery has become a central topic in urban analytics and human-centered urban design. However, most existing studies treat urban scenes as static and largely ignore the role of dynamic elements such as pedestrians and vehicles, raising concerns about potential bias in perception-based urban analysis. To address this issue, we propose a controlled framework that isolates the perceptual effects of dynamic elements by constructing paired street-view images with and without pedestrians and vehicles using semantic segmentation and generative inpainting. Based on 720 paired images from Dongguan, China, a perception experiment was conducted in which participants evaluated original and edited scenes (removal of pedestrians and vehicles) across six perceptual dimensions, including *wealth, safety, vibrancy, beauty, boredom, and depression*. The results indicate that removing dynamic elements leads to a consistent decrease in perceived vibrancy (30.97%), whereas changes in other dimensions are more moderate and heterogeneous. To further explore the underlying mechanisms, we trained 11 machine-learning models using multimodal visual features. Feature-importance analysis identifies lighting conditions, human presence, and depth variation as key factors driving perceptual change. At the individual level, 65% of participants exhibited significant vibrancy changes, compared with 35–50% for other dimensions, revealing notable inter-individual sensitivity; gender further showed a marginal moderating effect on safety perception. Beyond controlled experiments, the trained model was extended to a city-scale dataset covering 47,963 locations (191,852 images) to predict vibrancy changes after the removal of dynamic elements. The city-level results reveal that such perceptual changes are widespread and spatially structured, affecting 73.7% of locations and 32.1% of images, suggesting that urban perception assessments based solely on static imagery may substantially underestimate urban liveliness. Overall, this study highlights the critical role of dynamic elements in shaping urban perception and underscores the importance of accounting for transient urban features in large-scale perception-driven urban studies.

# 1. Introduction

Urban perception plays a central role in shaping how individuals experience and emotionally engage with their surroundings (Ito et al., 2024; Zhu et al., 2025). As cities become increasingly dense, diverse, and visually complex, understanding how people interpret and respond to urban environments has become a key concern for urban planning, landscape architecture, and public health (Cao et al., 2025). The way people perceive a city affects their emotional states, such as feeling safe, and also influences long-term behaviors, place attachment, and well-being (Wolch et al., 2014). Visual stimuli in the urban environment, including buildings, greenery, signage, and people, play a crucial role in shaping these perceptions (Greene & Oliva, 2009; Lindal & Hartig, 2013). Recent advances in computing and spatial analysis have made it possible to study these perceptions at scale, revealing how urban form and visual features correlate with feelings of wealth, safety, vibrancy, beauty, boredom, and depression (Dubey et al., 2016; Zhang et al., 2018). These developments have supported a growing body of data-driven research on urban perception, which not only enables empirical analysis of subjective spatial experience but also provides critical tools for improving the livability and sustainability of cities (Wu et al., 2022).

Building on this foundation, researchers have increasingly adopted methods based on street view imagery, semantic segmentation, and machine learning to quantify how specific urban elements shape emotional responses. For instance, studies have used deep learning to extract visual features from images and train models to predict human judgments across multiple perceptual dimensions (Dubey et al., 2016; Zhao et al., 2023; Zhang et al., 2023; Liang et al., 2024). Others have employed large-scale crowdsourced evaluations, such as the Place Pulse project (Salesses et al., 2013), a global collection of street images in 688 cities (Hou et al., 2024), or proposed adversarial learning frameworks to economically assess urban perceptions in Chinese cities (Yao et al., 2019). Some approaches also incorporate spatial syntax or visual hierarchy to enhance the interpretability of perception scores (Wang et al., 2022). More recently, researchers have integrated eye-tracking data to provide a more nuanced understanding of how people visually explore city scenes and which elements attract attention (Kiefer et al., 2024; Noland et al., 2017; Liao et al., 2019; Wu et al., 2022; Yang et al., 2024a; Kang et al., 2025). These methods capture visual saliency and attentional focus, enabling a richer model of how urban features such as cars, sidewalks, trees, or crowds influence perception.

Despite these advances, a key limitation persists: most existing studies rely on static street view images, which represent only a single moment in time and fail to capture the inherent temporal dynamics of urban environments (Yang et al., 2024b). In real-world settings, transient elements such as pedestrians, vehicles, and street activities constantly reshape how people experience the city (Vater, 2022; Qin et al., 2025). These moving objects not only contribute to the semantic meaning of a scene but also dominate the visual field and emotional interpretation. However, their role has largely been overlooked in existing computational models, which tend to treat the visual environment as static and uniform (Yao et al., 2025). A fundamental question therefore arises: how can we generate or manipulate dynamic urban scenes in a controlled way to better understand their perceptual effects? The recent progress of generative models provides a promising opportunity. Models such as Stable Diffusion (Rombach et al., 2022), DALL·E 3 (Betker et al., 2023), and LaMa (Suvorov et al., 2022) can synthesize high-quality urban scenes that include or exclude specific dynamic elements, such as pedestrians or vehicles. By leveraging these generative capabilities, researchers may create large-scale, semantically consistent datasets to test how dynamic features influence human perception of cities.

Based on the above considerations, this study tries to explore how the presence or absence of dynamic elements, specifically pedestrians and vehicles, affects people's perception of urban street scenes. We employ StabilityAI's Inpainting tool to remove dynamic elements from real street view images while preserving spatial and semantic coherence, and construct a paired dataset of 720 original and edited street scenes. Forty participants were then invited to rate subsets of these image pairs under randomized conditions, ensuring that they were unaware of the pairing design. Their evaluations covered six perceptual dimensions, including *wealth, safety, vibrancy, beauty, boredom, and depression*. We analyzed these results from both the image and individual levels to identify consistent perceptual differences. Machine learning models were further used to explore which visual characteristics of street scenes contribute most to these differences and to predict perceptual variations at the city scale.

## 2. Related works

Urban perception refers to the emotional experiences and cognitive evaluations individuals form in response to urban environments, encompassing feelings such as safety, vibrancy, beauty, and wealth (Dubey et al., 2016; Ordonez & Berg, 2014). Understanding these subjective experiences has become increasingly important for informing people-centered urban design, promoting mental health, and enhancing urban livability (Wang et al., 2019). Since studying urban perception through street imagery requires systematically acquiring data, applying analytical methods, and revealing the cognitive mechanisms underlying human responses, existing research has primarily developed along these three dimensions: (1) **data sources**, (2) **analytical approaches**, and (3) **mechanism exploration**.

**(1) Data source**

In terms of data, early studies relied heavily on traditional approaches such as surveys and interviews to capture individuals' perceptions of urban spaces (Kaplan & Kaplan, 1989; Nasar, 1990). Although valuable, these methods were often constrained by limited geographic coverage and subjective biases. The introduction of large-scale street view imagery, such as Google Street View and Baidu Street View, marked a major shift, enabling researchers to systematically collect visual representations of urban environments across diverse settings (Wang et al., 2025). Landmark projects such as Place Pulse (Salesses et al., 2013) compiled global datasets through crowdsourced pairwise comparisons, allowing researchers to link perceptions of safety, wealth, vibrancy, beauty, and activity to specific visual stimuli. Further expansions through Place Pulse 2.0 (Dubey et al., 2016) broadened the geographic scope and perceptual dimensions, providing a robust foundation for computational modeling. Other datasets, such as the UrbanPerception dataset (Ordonez & Berg, 2014), Global Streetscapes dataset (Hou et al., 2024), and a time-series street view imagery dataset for analyzing urban physical disorder in China (Ma et al., 2025), have further enriched the available resources for studying the visual and emotional perception of cities. In addition, Wang et al. (2025) evaluated the data quality and temporal availability of street view imagery from Google and Baidu. Taken together, these developments illustrate that as urban imagery data becomes increasingly abundant, the data source is also evolving from single-source to multi-source and from single-timestamp to multi-temporal characteristics.

**(2) Analytical approaches**

Building on these rich data resources, the analytical methods for studying urban perception have evolved substantially. Early computational efforts mainly employed traditional machine learning techniques, such as support vector machines (Naik et al., 2014) and random forests (Quercia et al., 2014), often relying on handcrafted visual features extracted from images. With the rise of deep learning, convolutional neural networks (CNNs) and related architectures were increasingly applied to directly learn perceptual patterns from raw visual inputs (Dubey et al., 2016; Porzi et al., 2015; Zhang et al., 2018; Fan et al., 2023; Ye et al., 2024). More recently, adversarial learning frameworks (Yao et al., 2019) and spatial syntax-based models (Wang et al., 2022) have been introduced to enhance prediction robustness and interpretability, reflecting a broader trend toward more sophisticated, end-to-end perceptual modeling. Alongside these developments, the emergence of foundation models has further broadened the methodological landscape, with large vision models increasingly being applied to tasks such as urban functional zone classification and poverty inference (Huang et al., 2024; Wu et al., 2025). These developments indicate that with the continuous advancements in computer vision and machine learning technologies, a wide range of analytical approaches have been introduced into the study of urban perception. Therefore, it should also be noted here that this review may not cover all existing approaches, as numerous similar methods have been proposed in recent years.

**(3) Mechanism exploration**

Beyond prediction accuracy, researchers have increasingly sought to understand the cognitive mechanisms underlying urban perception. Initial studies inferred attention and emotional responses indirectly through aggregated user ratings (Salesses et al., 2013; Yao et al., 2019). However, advances in physiological sensing technologies have enabled the direct capture of cognitive indicators such as eye movements (Noland et al., 2017; Liao et al., 2019; Wu et al., 2022; Yang et al., 2024a; Yang et al., 2024b) and, more recently, electroencephalography (EEG) signals (Hei et al., 2025; Yang et al., 2025). Eye-tracking experiments have demonstrated that attentional allocation to specific urban elements—such as pedestrians, vehicles, and vegetation—significantly influences perceived safety, vibrancy, and social presence. Complementarily, EEG studies have begun to reveal how different urban visual stimuli evoke distinct neural responses associated with emotional and cognitive processing. These multimodal insights mark an important step toward a deeper, cognitively grounded understanding of how urban environments are perceived.

In summary, research on urban perception has progressed from small-scale and subjective assessments to large-scale, data-driven analyses supported by street view imagery and deep learning. These advances have deepened our understanding of how visual and structural elements shape emotional responses to urban environments. However, most existing studies continue to rely on static representations of cities, failing to capture the perceptual effects of dynamic and transient elements such as pedestrians and vehicles. Recent advances in generative models now provide new opportunities to overcome this limitation. Building on this emerging paradigm, there remains a need to systematically analyze how dynamic elements, such as pedestrians and vehicles, influence urban perception.

**3. Methodology**

Figure 1 presents the experimental procedure of this study, which consists of two main processes. The first process focuses on stimulus preparation, in which large-scale street-view images are collected and paired stimuli are constructed by selectively removing dynamic elements, including pedestrians and vehicles, through semantic segmentation and generative inpainting (*Section* 3.1). The second process focuses on collecting perceptual evaluations from participants on both the original and edited scenes across multiple perceptual dimensions (*Section* 3.2). Together, these two processes form the basis for subsequent image-level and individual-level analyses.

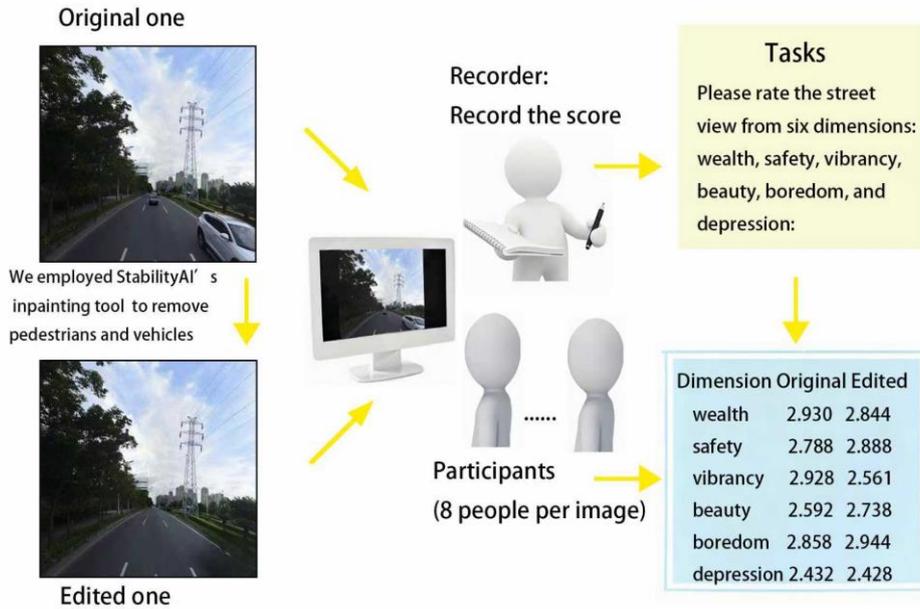

Figure 1. Experimental procedure

**3.1 Stimulus preparation**

To construct the experimental stimuli, we first extracted the road network of Dongguan, China, from OpenStreetMap (OSM). The location of Dongguan is shown in Figure 2. Dongguan is almost the fastest urbanizing city in southern China, which has received limited attention in street-view–based urban perception studies and was therefore selected as the study area.

At 50-meter intervals along the road network, panoramic street-view images were retrieved from Baidu Maps, yielding 47,963 locations collected between February 10 and February 14, 2025. To minimize distortions and artifacts introduced by the street-view capture vehicle, the camera perspective was adjusted by tilting the view upward by 15 degrees. From each panoramic image, four directional views (front, back, left, and right) were extracted to construct the initial street-view dataset, resulting in a total of 191,852 images. Subsequently, semantic segmentation was performed using DINOv2 (Oquab et al., 2023), a powerful foundation model released by Meta AI. The segmented visual elements include sky, trees, grass, buildings, roads, sidewalks, overpasses, vehicles, pedestrians, railings, walls, streetlights, and signposts. This categorization follows established practices in computational urban perception studies, which have demonstrated that these visual components play distinct roles in shaping human judgments of urban environments (Yang et al., 2024b; Cao et al., 2025). The overall distribution of the segmented visual elements is summarized in Table 1.

As shown in the table, we can observe that static elements such as sky (mean proportion: 32.53%), roads (21.57%), buildings (15.02%), and trees (16.72%) dominate the visual composition of urban scenes, reflecting the structural characteristics of the built environment. Notably, dynamic elements are also highly prevalent: vehicles appear in 78.85% of images, while pedestrians are present in 25.78% of images. These results indicate that dynamic elements such as vehicles and pedestrians also constitute a non-negligible component of street-view imagery and therefore warrant systematic investigation.

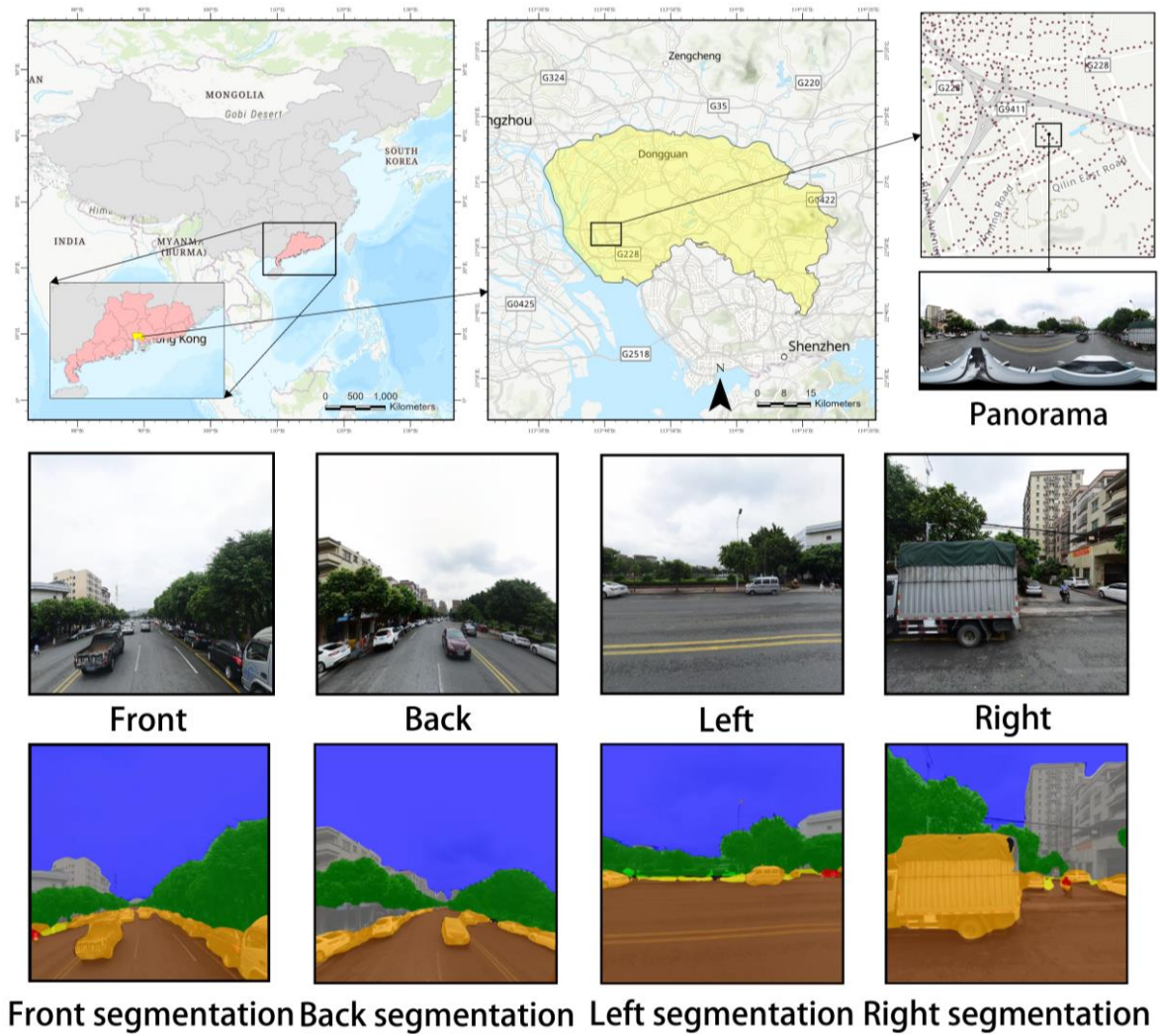

Figure 2．Data source and image segmentation.

Table 1. Results of visual element segment.

| Element | Frequency | Maximum | Minimum | Mean |
|---|---|---|---|---|
| Sky | 95.56% | 77.32% | 0.00% | 32.53% |
| Trees | 93.07% | 99.99% | 0.00% | 16.72% |
| Grass | 49.10% | 66.63% | 0.00% | 1.24% |
| Buildings | 89.97% | 100.00% | 0.00% | 15.02% |
| Roads | 97.17% | 53.45% | 0.00% | 21.57% |
| Sidewalks | 82.85% | 53.99% | 0.00% | 3.25% |
| Overpass | 6.64% | 70.02% | 0.00% | 0.23% |
| Vehicles | 78.85% | 98.29% | 0.00% | 3.51% |
| Persons | 25.78% | 34.25% | 0.00% | 0.08% |
| Railings | 8.72% | 33.62% | 0.00% | 0.07% |
| Walls | 53.68% | 100.00% | 0.00% | 1.43% |
| Streetlights | 37.02% | 3.32% | 0.00% | 0.03% |
| Signposts | 44.69% | 38.70% | 0.00% | 0.25% |

To systematically construct paired stimuli with and without dynamic elements, images in which dynamic elements (pedestrians and vehicles) accounted for more than 5% of the visual area were retained as candidate scenes, resulting in a street-view dataset enriched in dynamic content. For these candidate images, we employed StabilityAI's inpainting tool to remove pedestrians and vehicles. The inpainting procedure integrated the original street-view images, the corresponding segmentation masks of pedestrians and vehicles, and the textual prompt 'remove pedestrians and vehicles from the image'. This process generated paired versions of each scene: the original image containing dynamic elements and an edited image in which these elements were removed.

To ensure data quality, a consensus-based screening was conducted, and yielding 720 edited images in which the removal of dynamic elements appeared visually coherent and semantically consistent. These images were drawn from different view directions, including front (201 images), back (214 images), left (154 images), and right (151 images), ensuring a relatively balanced directional distribution. To facilitate the experimental procedure, the final dataset was further divided into 10 groups of original images (each containing 72 images) and 10 corresponding groups of edited images, with each edited group directly paired with its original counterpart, representing the same set of scenes with dynamic elements removed.

**3.2 Perceptual dimensions and participants**

(1) Perceptual dimensions

Following common practice in urban perception research (e.g., Salesses et al., 2013; Dubey et al., 2016; Wu et al., 2022), participants were asked to evaluate each scene along six perceptual dimensions: *wealth, safety, vibrancy, beauty, boredom, and depression*. These dimensions have been widely used to capture both positive and negative aspects of emotional and cognitive responses to urban environments. For each image, participants provided ratings on a five-point Likert scale (1 = very low, 5 = very high) for all six dimensions.

(2) Participants

A total of 40 participants (12 male, 28 female; aged from 20-32; all with normal or corrected-to-normal vision and no ocular diseases) were recruited for the experiment. Participants primarily consisted of undergraduate and graduate students, as well as faculty members in the field of Geographic Information Science, ensuring domain familiarity while maintaining diversity in age and experience. Each participant was randomly assigned two groups of images, which in fact represented the same set of urban scenes presented in two different conditions: one group contained the original street-view images with pedestrians and vehicles, and the other contained the corresponding edited images in which these dynamic elements had been removed. Each group consisted of 72 images, and every image was viewed and rated by at least eight different participants, thereby reducing individual-level noise and enhancing the robustness of the results. Each participant received a compensation of 80 RMB for completing the experiment.

## 3.3 Experimental procedure

All experiments were conducted in a well-lit laboratory room to ensure stable visual conditions. The procedure included three phases: an introductory session, a testing session, and the formal experimental session.

**(1) Introduction**. Participants were welcomed and provided with an overview of the study, and instructed on how to provide perceptual ratings. They were informed that the purpose of the experiment was to investigate urban perception, and that their task was to evaluate each image based on their subjective impressions across six perceptual dimensions (*wealth, safety, vibrancy, beauty, boredom, depression*). To maintain experimental control, participants were not informed that the same scenes would appear in both original and edited versions. A brief description of the experimental process was then provided.

**(2) Training**. Participants are asked to complete a short trial designed to ensure that they are fully familiar with the experimental procedure. Each participant first completed a block of five practice images, which followed exactly the same procedure as the formal experiment: participants observed each street-view image for 8 seconds, after which the screen switched to white, and they were asked to provide ratings on the six perceptual dimensions (*wealth, safety, vibrancy, beauty, boredom, depression*). After finishing this block, participants were asked whether they felt sufficiently familiar with the procedure. If they confirmed their familiarity, they proceeded to the formal experiment; if not, they were given an additional set of five practice images, again using the same sequence of viewing and rating. All practice trials were excluded from the final analysis.

(3) **Formal experiment**. The procedure is divided into participant tasks and implementation.

• **Tasks**. Each participant was asked to complete the rating of two groups of images, with each group containing 72 street-view scenes, as described in *Section* 3.1. One group consisted of the original images with pedestrians and vehicles, while the other group consisted of the corresponding edited images in which these dynamic elements had been removed. Participants evaluated each image on six perceptual dimensions (*wealth, safety, vibrancy, beauty, boredom, depression*) using a five-point Likert scale, based on their subjective impressions. To minimize memory effects, the two image groups were presented in separate sessions at least three days apart, following Ebbinghaus's forgetting curve (memory retention <30% after three days). Within each group, the order of image presentation was fully randomized.

• **Implementation.** Before the start of each formal session on a group of street-view images, participants completed a brief training block to re-familiarize themselves with the procedure. Each trial then followed a fixed sequence: a street-view image was displayed full-screen for 8 seconds. This was followed by a white screen, during which participants provided their ratings. The white-screen phase served both as the rating period and as an opportunity for participants to relax. After completing their ratings, participants pressed any key to proceed to the next trial.

## 4. Data analysis
### 4.1 Image-level analysis
#### 4.1.1 Descriptive statistics

According to the experimental design described in *Section* 3, each image (original vs edited) was evaluated by eight independent participants. We first averaged the ratings across participants to obtain an image-level score, and then conducted descriptive analyses based on these aggregated values. Table 2 summarizes the descriptive statistics of perceptual ratings across six dimensions for original and edited street-view images. As shown in the table, we can observe that wealth, safety, boredom, and depression showed relatively stable evaluations, with mean values changing only slightly between the two conditions (wealth: 2.930 to 2.844; safety: 2.788 to 2.888; boredom: 2.858 to 2.944; depression: 2.432 to 2.428). This suggests that the removal of dynamic elements had little impact on perceptions related to wealth, safety, or negative affect. In contrast, vibrancy showed a more pronounced difference, with the average rating decreasing from 2.928 in the original images to 2.561 in the edited ones, indicating that pedestrians and vehicles play a crucial role in conveying a sense of liveliness in urban scenes. By contrast, beauty exhibited the opposite trend: ratings increased from 2.592 to 2.738 after dynamic elements were removed, implying that the absence of pedestrians and vehicles may render scenes visually cleaner and more aesthetically pleasing.

In summary, the descriptive results suggest a differentiated effect of removing dynamic elements: vibrancy decreased while beauty increased, whereas other perceptual dimensions (wealth, safety, boredom, depression) remained relatively stable. These patterns provide initial evidence that the presence or absence of dynamic elements selectively shapes how urban scenes are perceived.

Table 2. Descriptive statistics of perceptual ratings for original and edited street-view images. Values with average rating changes within 0.1 are marked with a dash (-), while those exceeding 0.1 are indicated with ↑ or ↓ to represent the direction of change.

| Dimension | Original | | | | Edited | | | |
|---|---|---|---|---|---|---|---|---|
| | Min | Max | Ave | Std | Min | Max | Ave | Std |
| wealth | 1.500 | 4.750 | 2.930 | 0.621 | 1.250 | 4.750 | 2.844 (-) | 0.624 |
| safety | 1.125 | 4.500 | 2.788 | 0.481 | 1.500 | 4.375 | 2.888 (-) | 0.511 |
| vibrancy | 1.625 | 4.500 | 2.928 | 0.551 | 1.375 | 4.250 | 2.561 (↓) | 0.568 |
| beauty | 1.000 | 4.250 | 2.592 | 0.602 | 1.250 | 4.625 | 2.738 (↑) | 0.653 |
| boredom | 1.625 | 4.375 | 2.858 | 0.468 | 1.625 | 4.625 | 2.944 (-) | 0.474 |
| depression | 1.250 | 4.500 | 2.432 | 0.564 | 1.000 | 4.750 | 2.428 (-) | 0.609 |

**4.1.2 Perceptual differences at the image level**

To further examine whether the removal of dynamic elements (vehicles and pedestrians) produced statistically significant differences in perceptual evaluations, we conducted paired comparisons for each image based on the ratings provided by eight participants before and after editing (original vs. edited). For each perceptual dimension, we first applied the Shapiro–Wilk test, which is well-suited for small sample sizes, to assess whether the normality assumption was satisfied. When normality held, a paired-sample *t*-test was conducted; otherwise, the Wilcoxon test was used as a non-parametric alternative. The number of paired comparisons employed for each dimension is reported in Table 3. A significance threshold of $p < 0.1$ was adopted in this study due to the relatively small sample size available per image. Furthermore, employing a slightly more relaxed threshold also improves sensitivity to subtle yet potentially meaningful perceptual shifts that might otherwise be missed. Such use of trend-level criteria has precedent in perceptual and behavioral research, particularly when sample sizes are constrained (van de Schoot & Miočević, 2020). The significant results at the $p < 0.1$ level are shown in Table 3.

From the table, we can observe that almost all dimensions exhibited a certain proportion of significant differences between original and edited images, suggesting that the removal of dynamic elements did affect perception in a measurable way. The most pronounced effect was observed for vibrancy, with 30.97% of images showing significant differences and an overwhelming majority of 213 decreases vs 10 increases. This underscores the strong role of pedestrians and vehicles in conveying liveliness within urban scenes. The smallest proportion of significant results was observed for depression (8.61%), where increases and decreases were relatively balanced (25 vs. 37), suggesting heterogeneous responses across images. Between these two poles, other dimensions showed intermediate levels of sensitivity. Safety (14.44%) was most often associated with significant 82 increases vs. 22 decreases, implying that the removal of dynamic elements sometimes enhanced the perceived safety of scenes. Wealth (13.61%) was primarily associated with 74 decreases vs. 24 increases, suggesting that the absence of dynamic elements reduced perceived socioeconomic status. Beauty (13.47%) was strongly skewed toward 84 increases vs. 13 decreases, consistent with the notion that removing pedestrians and vehicles often improves aesthetic impressions. Boredom (12.78%) tended toward 64 increases vs. 28 decreases, indicating that edited scenes were sometimes experienced as more monotonous.

In summary, although the extent of significant differences varied across dimensions, the analysis demonstrates that a substantial subset of images consistently elicited perceptual changes after dynamic elements were removed. The effect was most pronounced for vibrancy (decrease), moderately present for safety (increase), beauty (increase), wealth (decrease), and boredom (increase), and weakest and mixed for depression.

Table 3. Number (*n*) and rate of image showing significant results (p < 0.1) across perceptual dimensions.

| Dimension | Number of paired comparisons | | Significant result $p<0.1$ | | | |
|---|---|---|---|---|---|---|
| | *t*-test | Wilcoxon test | $n(\uparrow)$ | $n(\downarrow)$ | $n$ (Total) | Rate (%) |
| wealth | 217 | 503 | 24 | 74 | 98 | 13.61 |
| safety | 303 | 417 | 82 | 22 | 104 | 14.44 |
| vibrancy | 374 | 346 | 10 | 213 | 223 | 30.97 |
| beauty | 345 | 375 | 84 | 13 | 97 | 13.47 |
| boredom | 369 | 351 | 64 | 28 | 92 | 12.78 |
| depression | 320 | 400 | 25 | 37 | 62 | 8.61 |

**4.1.3 Prediction of perceptual differences with multimodal features**

The preceding analyses revealed consistent perceptual differences before and after removing dynamic elements; the underlying question remains why such differences occur and which visual characteristics of urban scenes drive these perceptual shifts. To explore these mechanisms, we (1) extracted scene features, (2) built predictive models based on multimodal image features, and (3) analyzed which features contributed most to the observed perceptual differences.

**(1) Scene feature extraction**

Following established practices in computational urban perception research (e.g., Dubey et al., 2016; Zhang et al., 2018; Yang et al., 2024b; Cao et al., 2025), we thus extracted multimodal scene descriptors from three complementary perspectives—color, depth, and semantics—to characterize the structural and visual composition of each scene.

**Color features**: They describe the low-level visual tone and atmosphere of the scene. We calculated the distribution ratio of RGB components by equally dividing each color channel (R, G, and B) into 16 intervals (0-255) and computing the proportion of pixels within each range. For example, $R_{red\_1}$ represents the proportion of pixels whose red-channel values fall within the first interval (0-16), indicating the relative dominance of darker red tones in the image. In addition, we derived higher-order statistics such as color entropy and contrast, which have been shown to correlate with aesthetic and affective evaluations (Yang et al., 2024b).

**Depth features**: They reflect the three-dimensional organization of urban space and viewer immersion. Depth maps were generated using the MonoDepth2 model (a widely applied depth estimation model) (Godard et al., 2019), from which we derived mean depth ($Deep_{mean}$), standard deviation of depth ($Deep_{std}$), minimum depth ($Deep_{min}$), maximum depth ($Deep_{max}$), median depth ($Deep_{mode}$) to quantify the spatial openness and enclosure of each scene. These indicators help explain how perceived safety or vibrancy relates to spatial structure and visibility (Zhang et al., 2018).

**Semantic features**: They capture the composition of object categories within the scene. Based on DinoV2 segmentation (Oquab et al., 2023), we calculated the proportion of key elements (buildings, trees, sidewalks, roads, sky, etc.) and summarized them into 13 semantic categories. The relative dominance of these categories (e.g., greenery ratio, sky openness) provides a high-level description of urban form and land use context (Cao et al., 2025).

**(2) Classifier selection**

As outlined in *Section* 4.1.2, the sample sizes are imbalanced for perceptual differences on the six dimensions that appeared more frequently than others. This imbalance may lead to a bias toward the majority class in certain probability-based models. To address the issue, we chose the Synthetic Minority Over-sampling Technique (SMOTE) (https://imbalanced-learn.org/stable/) (Lemaître et al., 2017) to increase the number of samples in minority classes.

We then conducted a pilot study to evaluate 11 commonly used classifiers for identifying perceptual variation. All classifiers were tested using default hyperparameters and an 80/20 cross-validation method. Model performance was measured using mean accuracy (*acc*) and *F*1-score (*F*1), as defined in Equations (1) and (2).

$$acc = \frac{TP+TN}{TP+TN+FP+FN} \quad (1)$$

$$F1 = \frac{2TP}{2TP+FP+FN} \quad (2)$$

where *TP*, *TN*, *FP*, and *FN* represent true positive, true negative, false positive and false negative, respectively.

The results of the 11 classifiers for predicting perceptual differences with multimodal features are presented in Table 4. Among all models, both the Multilayer Perceptron (MLP) and XGBoost achieved consistently strong performance across the six perceptual dimensions. The MLP model reached the highest overall scores, with accuracies of 0.879 (*F*1 = 0.886) for wealth, 0.877 (*F*1 = 0.886) for safety, and 0.887 (*F*1 = 0.894) for boredom. And the XGBoost model performed competitively, particularly for depression (*acc* = 0.892, *F*1 = 0.896), boredom (*acc* = 0.876, *F*1 = 0.880), and beauty (*acc* = 0.856, *F*1 = 0.857), with stable results in all categories. Although the MLP slightly outperformed other models in overall metrics, XGBoost demonstrated comparable accuracy with higher computational efficiency and interpretability. Therefore, we selected XGBoost as the final model for subsequent feature importance analyses and the evaluation of the perceptual differences at the city-level.

Table 4. The performance of 11 classifiers for predicting perceptual differences with multimodal features.

| Classifier | Wealth | | Safety | | Vibrancy | | Beauty | | boredom | | Depression | |
|---|---|---|---|---|---|---|---|---|---|---|---|---|
| | *acc* | *F*1 | *acc* | *F*1 | *acc* | *F*1 | *acc* | *F*1 | *acc* | *F*1 | *acc* | *F*1 |
| GBoost | 0.842 | 0.848 | 0.826 | 0.833 | 0.696 | 0.700 | 0.848 | 0.851 | 0.869 | 0.872 | 0.863 | 0.868 |
| LightGBM | 0.825 | 0.834 | 0.810 | 0.812 | 0.700 | 0.694 | 0.843 | 0.846 | 0.858 | 0.863 | 0.853 | 0.857 |
| MLP | 0.879 | 0.886 | 0.877 | 0.886 | 0.708 | 0.718 | 0.862 | 0.869 | 0.887 | 0.894 | 0.874 | 0.886 |
| AdaBoost | 0.785 | 0.796 | 0.769 | 0.775 | 0.670 | 0.664 | 0.766 | 0.776 | 0.787 | 0.795 | 0.789 | 0.795 |
| DecisionTree | 0.680 | 0.690 | 0.719 | 0.740 | 0.573 | 0.620 | 0.716 | 0.745 | 0.720 | 0.733 | 0.778 | 0.790 |
| LDA | 0.672 | 0.679 | 0.662 | 0.682 | 0.588 | 0.608 | 0.677 | 0.702 | 0.664 | 0.700 | 0.696 | 0.719 |
| KNN | 0.627 | 0.716 | 0.674 | 0.750 | 0.559 | 0.646 | 0.669 | 0.738 | 0.704 | 0.763 | 0.685 | 0.757 |
| LR | 0.664 | 0.676 | 0.651 | 0.673 | 0.577 | 0.595 | 0.681 | 0.703 | 0.667 | 0.688 | 0.682 | 0.699 |
| SVC | 0.674 | 0.697 | 0.692 | 0.732 | 0.592 | 0.629 | 0.657 | 0.704 | 0.670 | 0.706 | 0.717 | 0.754 |
| XGBoost | 0.847 | 0.855 | 0.839 | 0.844 | 0.694 | 0.700 | 0.856 | 0.857 | 0.876 | 0.880 | 0.892 | 0.896 |
| RF | 0.765 | 0.780 | 0.808 | 0.813 | 0.646 | 0.659 | 0.775 | 0.785 | 0.790 | 0.797 | 0.810 | 0.816 |

**(3) Feature importance analysis**

Using the XGBoost classifier, we further analyzed the relative importance of multimodal features for predicting perceptual differences across six dimensions. We reported the first 10 features that contribute the most in Figure 3. As shown in the figure, we can have the following observations:

For wealth, the most influential predictors were $R_{light}$ (0.052) and $R_{person}$ (0.051), followed by $R_{blue\_1}$, $R_{tree}$, and $R_{red\_13}$. The strong role of lighting and pedestrian-related features indicates that well-lit environments and visible human activity are essential visual cues of prosperity in urban scenes. Illuminated signage, streetlights, and people on sidewalks often imply active commercial areas or economic vibrancy. Consequently, the removal of pedestrians and vehicles weakens these contextual signals, resulting in a reduced sense of affluence. The presence of vegetation ($R_{tree}$) also modestly contributed to perceived wealth, suggesting that landscaped greenery often coexists with higher-quality built environments.

For safety, $R_{light}$ again ranked highest (0.055), accompanied by $R_{sidewalk}$, $R_{wall}$, and $R_{green\_13}$. This combination reflects the visual structure of streetscapes where sufficient illumination, clear pedestrian pathways, and well-defined spatial boundaries contribute to feelings of security. Interestingly, $Deep_{min}$ also appeared among the top features, implying that greater spatial openness—after removing objects that previously occluded the view— may enhance perceived safety. In essence, eliminating vehicles and crowds may expose more continuous sightlines and reduce visual clutter, which participants tend to interpret as a safer, more predictable environment.

For vibrancy, $R_{wall}$, $R_{grass}$, and $R_{sidewalk}$ were the leading predictors, while $R_{person}$ ranked prominently as well. These elements correspond to areas of social interaction and human activity, such as open sidewalks, vegetated plazas, or visually active street façades. Their importance underscores that vibrancy arises from both spatial affordances and semantic cues of movement. The disappearance of pedestrians and vehicles removes these temporal and social indicators, producing scenes that appear static and less lively. The results thus quantitatively confirm that the decline in vibrancy ratings observed in *Section* 4.1.2 originates from the loss of dynamic human and vehicular components that typically animate urban space.

For beauty, $R_{sidewalk}$ (0.039) and $R_{wall}$ (0.035) were most influential, followed by $R_{light}$, $R_{blue\_3}$, and $R_{green\_13}$. These features represent organized spatial layouts and balanced chromatic compositions. The prominence of these static and structural cues indicates that, once dynamic elements are removed, the underlying spatial order becomes more visible and aesthetically pleasing. For example, removing vehicles and pedestrians often exposes architectural symmetry, coherent color gradients, and unobstructed spatial perspectives. The increase in perceived beauty is therefore not merely a subjective bias but a direct consequence of improved visual clarity and compositional harmony within the scene.

For boredom, $R_{person}$ (0.060) and $R_{light}$ (0.035) ranked highest, along with depth-related indicators such as $Deep_{std}$ and $Deep_{mode}$. Scenes with fewer people and less depth variability tend to evoke monotony and visual fatigue. After removing dynamic elements, the remaining urban spaces often appear more static, lacking motion, color contrast, or social cues that sustain engagement. Reduced depth variation further limits visual exploration, creating an impression of spatial uniformity. These results explain why participants frequently rated edited scenes as more boring, as the absence of movement and human presence diminishes both semantic and spatial dynamism.

For depression, $R_{light}$ (0.048) and $R_{person}$ (0.033) again played dominant roles, followed by $Deep_{max}$ and low-saturation color indicators such as $R_{red\_2}$ and $R_{blue\_4}$. The diminished contribution of light and human-related features corresponds to darker, emptier, and less inviting scenes, which may amplify feelings of desolation or confinement. In many edited images, the removal of pedestrians and vehicles left behind vacant spaces with uneven lighting and restricted depth contrast, conditions that psychologically align with isolation and emotional dullness. The role of depth features such as $Deep_{max}$ further suggests that limited spatial distance or enclosed perspectives contribute to depressive impressions by reducing the perceived openness of the urban environment.

In summary, the feature importance analysis demonstrates that the perceptual differences following the removal of dynamic elements are systematically linked to structural and semantic changes in urban imagery. Lighting and human presence consistently emerged as dominant factors shaping perceptions of wealth, vibrancy, and safety, while depth variation and color uniformity played stronger roles in feelings of boredom and depression. The absence of pedestrians and vehicles not only eliminates movement cues but also alters the visual balance of light, color, and depth, leading to measurable shifts in how people evaluate urban scenes.

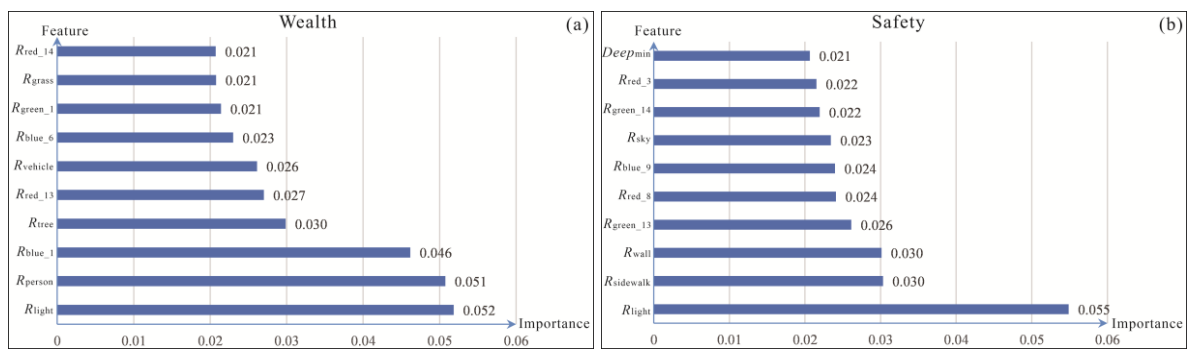

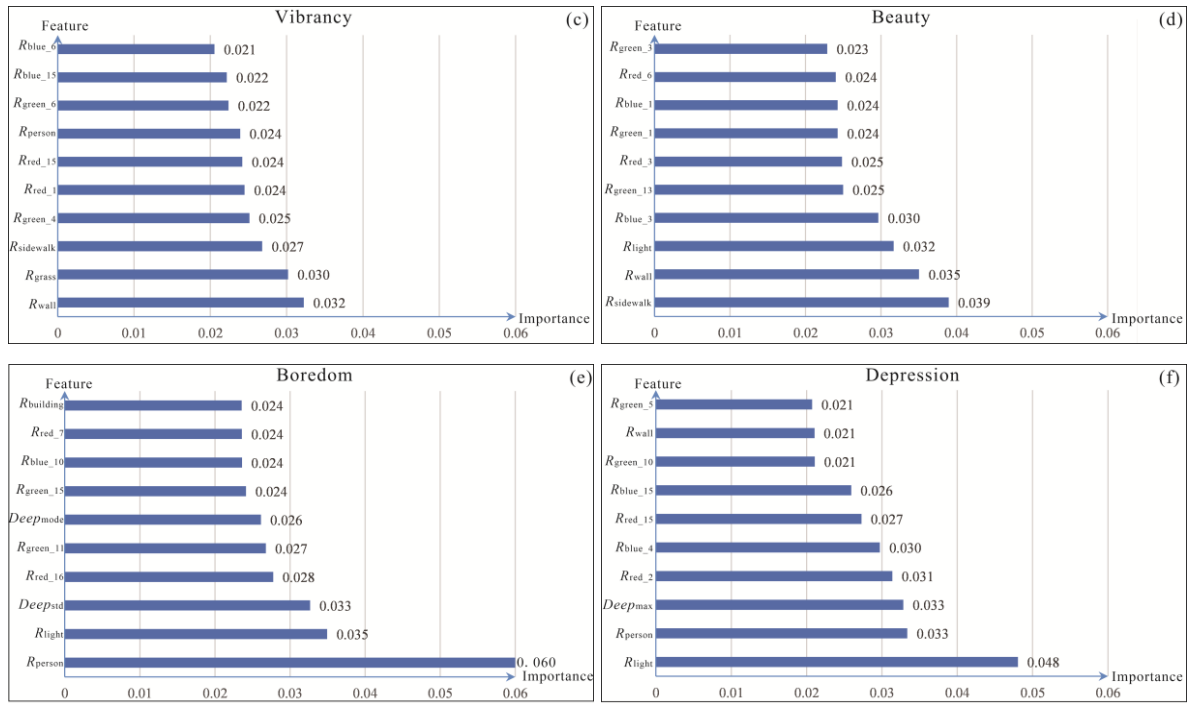

Figure 3. Feature importance in XGBoost for predicting perceptual differences (edited-original) across six perceptual dimensions.

## 4.2 Individual-level analysis

### 4.2.1 Descriptive statistics

According to the experimental design described in *Section* 3, we have collected the data of 40 participants, each of whom evaluated 72 street-view scenes in both their original and edited versions. For each participant, we averaged ratings across the 72 scenes within each condition (original vs. edited) and then calculated the difference between the two conditions across six perceptual dimensions. Figure 4 illustrates the distributions of these individual-level differences. As shown in the figure, we can observe that wealth, safety, boredom, and depression showed relatively stable evaluations, with most participants exhibiting only minor differences between the two conditions. This suggests that individual judgments of socioeconomic status, safety, and negative affect were largely unaffected by the removal of dynamic elements. In contrast, vibrancy demonstrated a consistent decline: nearly all participants rated edited scenes lower than original ones, reinforcing the critical role of pedestrians and vehicles in conveying a sense of liveliness. Conversely, beauty showed an upward shift for a majority of participants, implying that the removal of dynamic elements often enhanced the perceived visual quality of the scenes.

In summary, these individual-level observations are highly consistent with the image-level findings: vibrancy consistently decreased while beauty increased, whereas wealth, safety, boredom, and depression remained relatively stable. This parallel indicates that the effect of removing dynamic elements is not only evident across scenes but also systematically reflected in participants' evaluations.

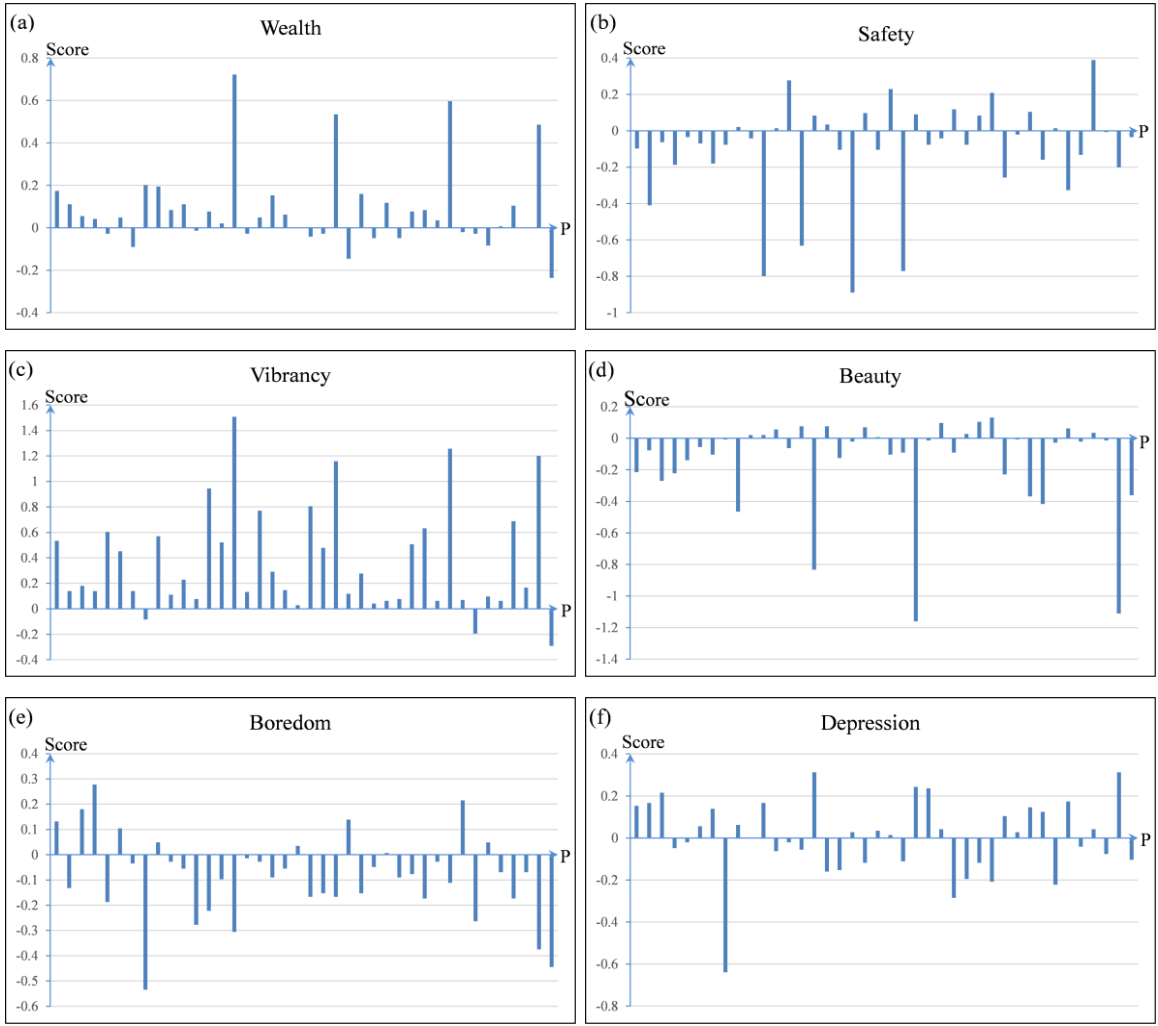

Figure 4. Distributions of individual-level rating differences (original-edited) across six perceptual dimensions.

**4.2.2 Perceptual differences at the individual-level**

To further examine whether the removal of dynamic elements (vehicles and pedestrians) produced statistically significant differences in perceptual evaluations at the individual level, we compared each participant's ratings for original and edited versions of the same scenes. For each perceptual dimension, the distribution of paired differences (edited – original) was first examined using the Shapiro–Wilk test, which is well-suited for small sample sizes. The results indicated that the normality assumption was not satisfied; therefore, the Wilcoxon signed-rank test was employed for all dimensions. The number of participants showing significant differences in their evaluations is reported in Table 5.

From the table, we can observe that almost all dimensions exhibited a certain proportion of significant differences, indicating that the removal of dynamic elements influenced perceptual judgments for a non-negligible subset of individuals. The most pronounced effect was observed for vibrancy, with 65.0% of participants showing significant differences and the vast majority exhibiting 24 decreases vs. 2 increases. This reinforces the critical role of pedestrians and vehicles in conveying a sense of liveliness. Boredom also displayed a relatively high proportion of significant results (50.0%), with most participants reporting 17 increases vs. 3 decreases, suggesting that edited scenes were often experienced as more monotonous. Safety (42.5%) showed more 11 increases vs. 6 decreases, implying that some participants perceived the removal of dynamic elements as enhancing the sense of safety. Depression (40.0%) was more balanced 7 increases vs. 9 decreases, indicating heterogeneous responses across individuals. At the lower end, wealth (35.0%) was dominated by 12 decreases 12 vs. 2 increases, suggesting that socioeconomic impressions weakened when dynamic elements were removed, while beauty (35.0%) was dominated by 13 increases vs. 1 decrease, implying that the absence of pedestrians and vehicles often improved aesthetic judgments.

In summary, although the extent of significant differences varied across participants, the individual-level analysis demonstrates that perceptual changes were consistently present. The strongest and most consistent effect was the decrease in vibrancy, followed by increases in boredom, safety, and beauty, while wealth tended to decrease, and depression showed mixed responses.

Table 5. Number ($n$) and rate of participants showing significant differences in their perceptions ($p < 0.05$).

| Dimension | $n$ (↑) | $n$ (↓) | $n$ (Total) | Rate (%) |
|---|---|---|---|---|
| wealth | 2 | 12 | 14 | 35.0 |
| safety | 11 | 6 | 17 | 42.5 |
| vibrancy | 2 | 24 | 26 | 65.0 |
| beauty | 13 | 1 | 14 | 35.0 |
| boredom | 17 | 3 | 20 | 50.0 |
| depression | 7 | 9 | 16 | 40.0 |

#### 4.2.3 Analysis of moderating factors

To further examine whether perceptual differences varied across participant characteristics, we analyzed potential moderating effects of gender. Specifically, we aimed to determine whether male and female participants differed in their sensitivity to perceptual changes caused by the removal of dynamic elements. Given that both gender and significance of perceptual difference are categorical variables, the Pearson chi-square test was applied to assess whether the distribution of significant results differed between the two gender groups. This non-parametric method is suitable for identifying associations between categorical factors without assuming data normality (Sloane & Morgan, 1996). The results are shown in

Table 6. From the table, we can observe that only safety showed a marginally significant gender difference ($\chi^2$ = 2.771, $p$ = 0.096*), indicating that female participants were more likely to exhibit significant perceptual changes in safety evaluations after the removal of pedestrians and vehicles. This suggests that women may be more sensitive to spatial openness and social cues influencing perceived security. For other perceptual dimensions—including wealth ($p$ = 0.193), vibrancy ($p$ = 0.285), beauty ($p$ = 0.193), boredom ($p$ = 0.168), and depression ($p$ = 0.121)—no significant gender differences were detected. Nevertheless, the general tendency showed that female participants had slightly higher rates of perceptual sensitivity overall.

In summary, the chi-square results suggest that gender exerts a selective moderating effect on urban perception, particularly for safety-related evaluations, while the perceptual impact of dynamic element removal remains largely consistent across genders.

Table 6. Results of the Pearson chi-square test.

| Dimension | $\chi^2$ | $p$ |
|---|---|---|
| wealth | 1.695 | 0.193 |
| safety | 2.771 | 0.096* |
| vibrancy | 1.143 | 0.285 |
| beauty | 1.695 | 0.193 |
| boredom | 1.905 | 0.168 |
| depression | 2.401 | 0.121 |

**Note**: ***, **, and * represent the 1%, 5%, and 10% significance levels, respectively.

## 5. Evaluate the perceptual differences in city-level

To examine whether the perceptual impact of dynamic elements persists at the city scale, we further conducted a city-level evaluation focusing on vibrancy, the perceptual dimension that exhibited the most substantial changes in both image-level and individual-level analyses in *Section* 4. Using the XGBoost model introduced in *Section* 4.1.3, we extended the prediction to the entire city of Dongguan. Specifically, perceptual differences were predicted for 47,963 street-view locations and 191,852 images, with four directional views (front, back, left, and right) available for each location.

The results indicate that removing dynamic elements leads to widespread changes in perceived vibrancy at the city scale. Among all images, 61642 images exhibited a predicted change in vibrancy after dynamic elements were removed, accounting for 32.1% of the total street-view images. When analyzed at the location level, 35358 locations have images that appear to be perceived differently on vibrancy, accounting for 73.7%. In particular, 16,752 locations showed perceptual changes in one view, 12,256 locations in two views, 5,422 locations in three views, and 1,028 locations in all four views, corresponding to 34.9%, 25.6%, 11.3%, and 2.1% of all locations, respectively. This distribution suggests that perceptual changes are not isolated to a small subset of scenes, but occur consistently across multiple perspectives for a substantial number of urban locations.

We also then visualized the distribution of vibrancy differences across Dongguan, as shown in Figure 5. The study area was divided into regular spatial grids. Each grid cell contains approximately 18.2 street-view locations on average. For each grid, we counted the number of images within the cell that exhibited a significant predicted change in vibrancy after dynamic elements were removed, and summarized this count as a grid-level indicator. As shown in the left panel, grids with higher counts of perceptual change are unevenly distributed across the city, forming clear spatial clusters rather than random noise. Areas with intensive road networks and active urban functions tend to exhibit higher concentrations of vibrancy change, whereas regions with more homogeneous or less active streetscapes show relatively stable perceptual outcomes. This pattern indicates that the perceptual impact of dynamic elements is spatially structured and closely related to urban activity intensity. While in the right panel, we provide four qualitative illustrations of these results through typical panoramic street-view examples. The top two cases (#40410 and #16923) show scenes whose predicted vibrancy values remain unchanged after dynamic elements are removed (Value = 0), reflecting environments where static physical structures dominate perceptual impressions. In contrast, the bottom two cases (#28377 and #19144) exhibit substantial changes in vibrancy (Value = 93 and Value = 94, respectively). In these scenes, pedestrians and vehicles constitute salient visual cues of social activity; once removed, the scenes become perceptually quieter and less lively despite identical underlying physical layouts. These results demonstrate that perceptual changes induced by dynamic element removal are both spatially systematic and visually interpretable. These results further support the conclusion that city-scale evaluations of urban vibrancy that ignore dynamic elements may fail to capture substantial and spatially heterogeneous aspects of human urban perception.

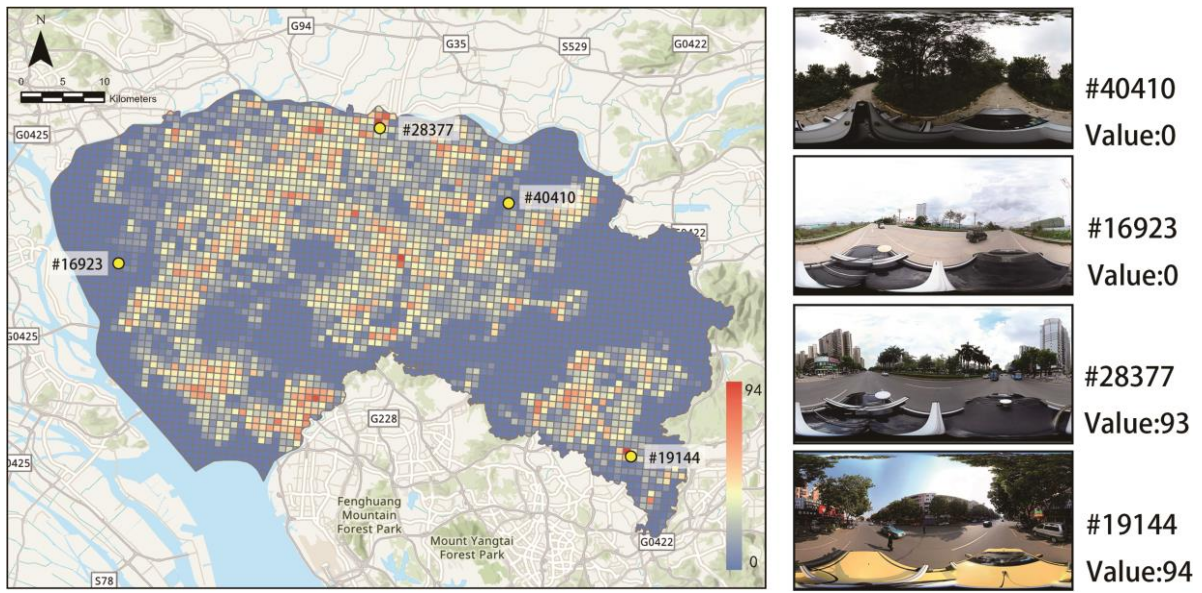

Figure 5. City-level spatial distribution of predicted vibrancy changes after removing dynamic elements and representative street-view examples.

## 6. Discussion

### 6.1 The role of dynamic elements in urban perception

Our findings highlight the differentiated impact of dynamic elements, such as pedestrians and vehicles, on urban perception. Among the six perceptual dimensions, vibrancy was most strongly influenced, with 30.97% of images and 65.0% of participants showing significant decreases after dynamic elements were removed. This underscores the crucial role of human presence and mobility in conveying liveliness within urban scenes. In contrast, beauty tended to increase (13.47% of images and 35.0% of participants, the vast majority in the upward direction), suggesting that dynamic elements may introduce visual clutter or distraction, while their absence can render environments cleaner and more aesthetically appealing. For other dimensions, the patterns were more modest but still noteworthy: safety showed increases in a subset of cases (14.44% of images, mostly increases; 42.5% of participants, with 11 increases vs. 6 decreases), wealth tended to decrease (13.61% of images, mostly decreases; 35.0% of participants, 12 decreases vs. 2 increases), boredom showed limited but noticeable increases (12.78% of images and 50.0% of participants, mostly increases), and depression displayed the smallest and most mixed effects (8.61% of images; 40.0% of participants, with nearly balanced increases and decreases). Collectively, these findings demonstrate that dynamic elements selectively shape how urban spaces are perceived, amplifying impressions of liveliness, often diminishing perceptions of socioeconomic status, while at times enhancing aesthetic evaluations and altering judgments of safety and boredom.

The differentiated patterns can be interpreted in light of the distinct roles that dynamic elements play in shaping perceptual impressions. Pedestrians and vehicles serve as salient markers of social activity and urban vitality, which explains why their absence significantly diminishes perceptions of vibrancy. At the same time, the removal of such elements often produces a visually cleaner environment, which may enhance aesthetic appeal and thus elevate ratings of beauty. Conversely, wealth and safety appear to depend more on static environmental cues such as building quality, greenery, or streetscape design, but the observed tendency for wealth to decrease and safety to increase indicates that dynamic elements still provide subtle cues to socioeconomic and security evaluations. These results extend previous street-view perception studies, which have predominantly focused on static physical attributes, by showing that dynamic elements constitute an additional layer of perceptual information that should not be neglected.

**6.2 Consistency between individual- and image-level results**

A second important finding is the high degree of consistency observed between image-level and individual-level analyses. At both levels, vibrancy reliably decreased following the removal of dynamic elements, whereas beauty generally increased. Other dimensions showed similar moderate patterns across levels: wealth decreased, safety and boredom increased, and depression remained mixed with no clear direction. This convergence indicates that the observed patterns are not merely artifacts of aggregation across images, but also systematically reflected in the judgments of individual participants. Such alignment strengthens the robustness of our conclusions and highlights the pervasive influence of dynamic elements across different units of analysis.

At the same time, our analyses also reveal meaningful individual differences, particularly for safety and depression, where a considerable share of participants showed significant changes (42.5% and 40.0% respectively) even though image-level effects were modest. While aggregate results suggest relative stability, these findings imply that personal experiences, attentional preferences, or cultural contexts may moderate sensitivity to dynamic elements. This points to a promising avenue for future research: expanding the participant pool to more diverse populations and cross-cultural contexts could help disentangle the conditions under which individual variability emerges. By integrating both the consistency of group-level patterns and the heterogeneity of individual responses, our study underscores the complex but systematic ways in which dynamic elements shape urban perception.

**6.3 Implications for urban design and involved research**

Our study clearly demonstrates that urban perception differs depending on whether dynamic elements such as pedestrians and vehicles are present in street-view scenes. This evidence calls for a reconsideration of how current applications and research incorporate perceptual dimensions, as existing research has been conducted almost exclusively on static street views, without accounting for these perceptual differences. In particular, we will discuss the implications from two perspectives: urban planning and design, and the related research on data-driven urban analysis.

**(1) Urban planning implications**

Our findings demonstrate that perceptions of vibrancy decrease when dynamic elements are removed, while perceptions of beauty often increase. These differentiated effects may provide actionable insights for urban design and planning. For example, if the objective is to foster a lively and vibrant atmosphere, cities may consider strengthening the visibility of dynamic elements by encouraging street activities, promoting pedestrian-friendly environments, and supporting spaces for social interaction. Conversely, if the goal is to highlight aesthetic order or cleanliness, measures such as traffic management, reducing irregular vehicle parking, or limiting overcrowded pedestrian zones could help minimize visual clutter and enhance perceived beauty. In this way, dynamic elements can be deliberately leveraged or mitigated to align with the intended perceptual goals of urban spaces.

**(2) Data-driven urban analysis**

Our findings demonstrate that perceptual evaluations vary systematically depending on the presence or absence of dynamic elements, which implies that these factors cannot be overlooked in data-driven approaches to urban analysis. Traditional urban perception studies and machine learning models have relied almost exclusively on static street-view imagery, thereby neglecting the perceptual influence of pedestrians and vehicles. If the goal is to capture the full spectrum of human perception, dynamic elements should be explicitly considered. In some cases, they could be systematically included or excluded to test their perceptual effects in smart city applications, virtual reality environments, or simulation platforms. This process can be operationalized through the trained model described in *Section* 4.1.3, with the potential for greater reliability if more perceptual data are incorporated. In other cases, especially when the research focus is on the structural or environmental features of the built environment, removing dynamic elements may help achieve more consistent evaluations by reducing noise from transient social activities. This dual strategy—selective inclusion or suppression—may support 'perception-oriented' urban design and opens new pathways for integrating perceptual science with digital urban analytics.

**6.4 Limitations and future works**

Despite the contributions in this research, several limitations should be acknowledged. First, the participant sample was relatively small (40 individuals) and predominantly composed of students and faculty in Geographic Information Science, which may constrain the generalizability of the results to broader populations. Second, only the main dynamic elements in the urban scene (pedestrians and vehicles) were examined, while other potentially influential factors, such as cyclists, animals, or temporary street facilities, were not included. Third, although the inpainting process used to remove dynamic elements was carefully validated, subtle artifacts may still have influenced participant judgments.

Future research should address these limitations by incorporating more diverse participant groups, expanding the scope of dynamic elements considered, and employing cross-cultural designs. In addition, integrating computational models with perceptual experiments could further illuminate how dynamic elements interact with static urban features to shape perceptions.

**7. Conclusion**

To investigate the role of dynamic elements in urban perception from street-view imagery and assess the bias introduced by treating urban scenes as static. We proposed a controlled framework that constructs paired street-view images with and without pedestrians and vehicles, enabling direct comparisons of perceptual responses. Perception experiments show that removing dynamic elements leads to a consistent decline in perceived vibrancy, while effects on other perceptual dimensions are more moderate and heterogeneous. The results further reveal notable inter-individual differences, indicating that urban perception is jointly shaped by environmental cues and individual sensitivity. By integrating multimodal visual features and machine-learning models, we demonstrate that dynamic elements influence urban perception through both direct visual saliency and interactions with lighting and spatial structure. Extending the analysis to a city-scale dataset covering nearly 48,000 locations shows that perceptual changes induced by dynamic element removal are widespread and spatially structured. These findings suggest that perception-based urban assessments relying solely on static imagery may systematically underestimate urban liveliness. At the individual level, participants varied in their sensitivity to the presence of dynamic elements, indicating that urban perception is shaped not only by visual content but also by individual perceptual characteristics. Overall, this work highlights the necessity of incorporating dynamic elements into perception-driven urban analysis and provides a scalable methodological pathway for moving beyond static representations of cities.

Despite these findings, this study has several limitations. The perception experiment was conducted in a single city with a relatively modest number of participants, which may limit the generalizability of individual-level responses. In addition, the analysis focused on pedestrians and vehicles as dynamic elements, while other transient urban factors were not considered. Future work will extend this framework to multiple cities and cultural contexts, incorporate a broader range of dynamic urban features, and further explore temporal variations in urban perception.

**Acknowledgments**

The authors would like to thank the editors and anonymous reviewers for their useful comments on the manuscript.

**Disclosure statement**

No potential conflict of interest was reported by the author(s).

**Funding**

This work was supported by grants from the National Natural Science Foundation of China (No. 42501551, 42371455, 42171438), Tobii China Innovation Initiative Project (TPI250407CN).